\def\la{\hbox{{\lower -2.5pt\hbox{$<$}}\hskip -8pt\raise
-2.5pt\hbox{$\sim$}}}
\def\ga{\hbox{{\lower -2.5pt\hbox{$>$}}\hskip -8pt\raise
-2.5pt\hbox{$\sim$}}}
\def\ltsima{$\; \buildrel < \over \sim \;$}
\def\simlt{\lower.5ex\hbox{\ltsima}}
\def\gtsima{$\; \buildrel > \over \sim \;$}
\def\simgt{\lower.5ex\hbox{\gtsima}}

\documentstyle[epsfig]{elsart}

\begin{document}
\begin{frontmatter}
\title{A semi-analytical approach to non-linear shock acceleration}
\author[Fermi]{Pasquale Blasi\thanksref{corr}} 
\address[Fermi]{NASA/Fermilab Theoretical Astrophysics Group,\\
Fermi National Accelerator Laboratory, Box 500, Batavia, IL 60510-0500}
\thanks[corr]{E-mail: blasi@fnal.gov}

\begin{abstract}
Shocks in astrophysical fluids can generate suprathermal 
particles by first order (or diffusive) Fermi acceleration. In the test
particle regime there is a simple relation between the spectrum of the
accelerated particles and the jump conditions at the shock. This simple
picture becomes complicated when the pressure of the accelerated particles
becomes comparable with the pressure of the shocked fluid, so that the 
non-linear backreaction of the particles becomes non negligible and the
spectrum is affected in a substantial way. Though only numerical simulations
can provide a fully self-consistent approach, a more direct and easily
applicable method would be very useful, and would allow to take into account 
the non-linear effects in particle acceleration in those cases in which 
they are important and still neglected.

We present here a simple semi-analytical model that deals with these 
non-linear effects in a quantitative way. This new method, while
compatible with the previous simplified results, also provides a satisfactory 
description of the results of numerical simulations of shock acceleration. 
\end{abstract}

\end{frontmatter}

\section{Introduction}
 
Diffusive shock acceleration is thought to be responsible for acceleration
of cosmic rays in several astrophysical environments. Most of the observational
evidence for this mechanism, also known as first order Fermi acceleration, 
has been provided by studies of heliospheric shocks, but there are indirect
lines of evidence that acceleration occurs at other shocks. A particularly
impressive example was provided a few years ago by the observation of 
gamma ray emission from the supernova remnant SN1006 \cite{sn1006}. These 
observations could be interpreted as inverse Compton emission of very high 
energy electrons, accelerated at the shock on the rim of SN1006, though
other radiation processes may contribute \cite{aa99}.

Shock acceleration has been studied carefully and a vast literature
exists on the topic. Some recent excellent reviews have been written
\cite{drury83,be87,bk88,je91}. Some of the more problematic aspects of the 
theory of particle acceleration at astrophysical shocks have been 
understood, while others are still subject of investigation. 

One of the problems that are harder to face is the problem of the injection
of particles in the acceleration region. Only particles with a Larmor radius
larger than the thickness of the shock are actually able to {\it feel} the
discontinuity at the shock. The shock thickness is of the order of the
Larmor radius of thermal protons, so that only a small fraction of the
particles can be accelerated.

The calculation of the injection efficiency is quite problematic, for several 
reasons: first, the distribution function of thermal particles is steeply
decreasing with momentum, so that the number of accelerated particles
changes wildly with changing injection momentum. Moreover the distribution 
of the particles in the shock frame is strongly anisotropic at these low
momenta, which adds to the difficulty of obtaining a straight analytical
answer. 
Second, the injection of particles from the thermal distribution and 
their subsequent diffusive transport is thought to be due to the scattering
against plasma waves, which are likely to be excited by the particles
themselves, which makes the problem intrinsically non-linear.
This non-linearity is exacerbated by the backreaction of the particles
on the structure of the shocked fluid \cite{bell87}.
The only way to have a complete quantitative picture of the problem of 
shock acceleration is to use numerical simulations 
\cite{elli90,je91,ebj95,ebj96,kj97}.

All these complicated effects, which seem to be 
important in several astrophysical situations, are nevertheless often 
neglected, mainly because of the lack of an approach that allows to
take them into account without the use of complicated numerical 
simulations which are usually of restricted use. 
As a consequence, in most of the applications of the diffusive shock 
acceleration to astrophysical situations, the assumption of test 
particles is adopted, even in those cases where this approximation 
works poorly. 
Numerical simulations show however that even when the fraction of particles
injected from the plasma is relatively small, the energy channelled into
these few particles can be close to the kinetic energy of the unshocked
fluid, making the test particle approach unsuitable. The most visible 
effect is on the spectrum of the accelerated particles, which shows
a peculiar flattening at the highest energies, due to the backreaction 
of accelerated particles on the fluid. The consequences on the spectra
of secondary particles and radiation processes are clear.

The need to have a theoretical understanding of the non-linear effects
in particle acceleration fueled many efforts in finding some
{\it effective} though simplified picture of the problem. 
The structure of shocked fluids with a backreaction of accelerated
particles was investigated in \cite{dr_v80,dr_v81,dr_ax_su82,ax_l_mk82,ddv94}
in a {\it fluid} approach. The thermodynamic quantities were calculated
including the effects of cosmic rays, but the approach did not provide
information on the spectral shape of the accelerated particles.

In Ref. \cite{blandford80} a perturbative approach was adopted, in which 
the pressure of accelerated particles was treated as a small perturbation.
By construction this method provides an answer only for weakly modified
shocks.

An alternative approach was proposed in \cite{eich84a,eich84b,eich85,elleich85}.
This approach is based on the assumption that the diffusion of the particles
is sufficiently energy dependent that different parts of the fluid are
affected by particles with different average energies. The way the calculations
are carried out implies a sort of separate solution of the transport equation
for subrelativistic and relativistic particles, so that the two spectra must 
be connected at $p\sim mc$ {\it a posteriori}. 

Recently, in \cite{berezhko94,berezhko95,berezhko96}, the effects of the 
non-linear backreaction of accelerated particles on the maximum achievable
energy were investigated, together with the effects of geometry. The solution 
of the transport equation was written in \cite{berezhko96} in an implicit 
form, and then expanded in terms of the unperturbed (linear) solution. 

Recently, some analytical solutions were also presented for the non-linear
shock acceleration, in the particular case of B\"{o}hm diffusion coefficient
\cite{malkov1,malkov2}. 

The need for a {\it practical} solution of the acceleration problem in the 
non-linear regime was recognized in \cite{simple}, where a simple analytical
approximation of the non-linear spectra was presented. In this model the 
spectrum of the accelerated particles was assumed to consist of a broken 
power law, with three slopes characterizing the low, intermediate and high 
energy regimes. The basic features of the spectra derived from numerical 
simulations were reproduced with this method. 

In the present paper we propose an approach that puts together some
of the elements introduced in \cite{eich84a}, \cite{berezhko96}
and \cite{simple} and provides a semi-analytical solution for the spectrum 
of accelerated particles and for the structure of the shocked fluid.
The method proposed is of simple use, can be adapted to several 
situations and provides results in very good agreement with numerical 
simulations, and with simplified models as that in \cite{simple}.

The paper is structured as follows: in \S2 we describe the general 
problem of linear and non-linear shock acceleration; in \S3 we 
explain in detail our approach to non-linear effects in shock acceleration.
In \S4 we discuss the results of our model and compare them with 
the predictions of the model in Ref. \cite{simple} and with the results
of some numerical simulations. Our discussion and conclusions are presented 
in \S5.

\section{Shock acceleration: linear and non-linear}

In this section we discuss the basic elements of shock acceleration 
and introduce our approach to the description of the non-linear effects due to 
the backreaction of the accelerated particles on the shocked fluid. 

For simplicity we limit ourselves to the case of one-dimensional shocks,
but the introduction of different geometrical effects is relatively simple, 
and in fact many of our conclusions will be not affected by geometry. 
The non-linear effects are restricted to the mutual action of the 
shocked fluid and the accelerated particles. In other words, the
present work does not include self consistently the production and the 
absorption of plasma waves by the accelerated particles. This simplification
is common to the approaches in \cite{berezhko96} and \cite{simple}.

The equation that describes the diffusive transport of particles 
in one dimension is
\begin{equation}
\frac{\partial f}{\partial t} = \frac{\partial}{\partial x}
\left[ D  \frac{\partial}{\partial x} f \right] + 
u  \frac{\partial f}{\partial x} + \frac{1}{3} 
 \frac{\partial u}{\partial x} p \frac{\partial f}{\partial p}
+ Q,
\label{eq:trans}
\end{equation}
where $f(x,p)$ is the distribution function, $u$ is the fluid 
velocity and $D$ is the diffusion coefficient. The injection of
particles is assumed to occur only immediately upstream of
the shock, so we write the source function as $Q=Q_0(p) \delta(x)$,
where $x=0$ corresponds to the position of the shock front.
For monoenergetic injection, the function $Q_0(p)$ has the
following form:
\begin{equation}
Q_0(p)=\frac{N_{inj} u_1}{4\pi p_{inj}^2} \delta(p-p_{inj}),
\label{eq:inj}
\end{equation}
where $p_{inj}$ is the injection momentum and $u_1$ is the fluid velocity 
immediately upstream ($u_1=u(0^+)$). $N_{inj}$ is the number density of
particles injected at the shock, parametrized here as $N_{inj}=\eta N_{gas,1}$,
where $N_{gas,1}$ is the gas density at $x=0^+$.
The boundary condition at the shock reads:
\begin{equation}
\frac{u_1-u_2}{3} p  \frac{\partial f_0}{\partial p} =
\left(D  \frac{\partial f}{\partial x}\right)_1 -
\left(D  \frac{\partial f}{\partial x}\right)_2 + Q_0(p),
\label{eq:boundary}
\end{equation}
where $u_2$ is the fluid velocity downstream.
Here we called $f_0=f(0,p)$ the distribution function at the shock 
position.

A useful way of handling eq. (\ref{eq:trans}) was suggested in Ref.
\cite{berezhko96} (a similar approach was also adopted in Ref. 
\cite{pell}), and consists of integrating this equation 
in the variable $x$ from $x=0^+$ (upstream) to $x=+\infty$
(far upstream). After some simple algebraic steps, in which we make
use of eq. (\ref{eq:boundary}), we obtain the following equation:
\begin{equation}
p\left(\frac{\partial f_0}{\partial p}\right) = 
-\frac{3}{u_p-u_2} \left\{ f_0 \left[ u_p + \frac{1}{3}
\frac{du_p}{d\ln p}\right] - Q_0(p)\right\},
\label{eq:spec}
\end{equation}
where the stationarity assumption was adopted and we assumed 
$(D\partial f/\partial x)_2=0$ . We have introduced the quantity:
\begin{equation}
u_p = u_1 + \frac{1}{f_0(p)} \int_0^{\infty} dx
\left(\frac{du}{dx}\right) f(x,p),
\label{eq:up}
\end{equation}
and again we called $u_1$ and $u_2$ the fluid velocities at $x=0^+$ 
and $x=0^-$ respectively. With this formalism the compression
factor at the shock is $R_{sub}=u_1/u_2$. 
The function $u_p$, at each momentum $p$ has the meaning of 
average fluid velocity felt by a particle with momentum $p$
while diffusing upstream. Since the diffusion is in 
general $p$-dependent, particles with different energies
will {\it feel} a different compression coefficient, and the
correspondent local slope of the spectrum will be 
$p$-dependent. Note that, according to eq. (\ref{eq:up}), the 
velocity $u_p$ must be a monotonically increasing function of $p$.

The function $u_p$ describes the mutual interaction between the
accelerated particles and the fluid. In other words,
if we find the way of determining the function $u_p$, as we 
show later, we also determine the spectrum of the accelerated 
particles.

Eq. (\ref{eq:up}) is clearly an implicit definition of $u_p$,
meaning that $u_p$ depends on the unknown
function $f$. However, eq. (\ref{eq:up}) allows us to extract 
important physical information. For the monoenergetic
injection in eq. (\ref{eq:inj}) the solution can be 
implicitly written as
\begin{equation}
f_0(p)=\frac{N_{inj} q_s}{4\pi p_{inj}^3} \exp
\left\{-\int_{p_{inj}}^p \frac{dp}{p} \left[
\frac{3 u_p}{u_p-u_2} + \frac{1}{u_p-u_2}\frac{du_p}{d\ln p}
\right]\right\},
\label{eq:implicit}
\end{equation}
where we put (for $\gamma_g=5/3$):
\begin{equation}
q_s=\frac{3R_{sub}}{R_{sub}-1}.
\end{equation}
Eq. (\ref{eq:implicit}) tells us that the spectrum of 
accelerated particles has a local slope given by
\begin{equation}
Q(p)=-\frac{3 u_p}{u_p-u_2} - \frac{1}{u_p-u_2}\frac{du_p}{d\ln p}.
\label{eq:slope}
\end{equation}
The problem of determining the spectrum of accelerated particles
would then be solved if the relation between $u_p$ and $p$ 
is found \footnote{In order to determine the spatial distribution of fluid
velocity it is necessary to specify the exact diffusion coefficient as 
function of $x$ and $p$. In this paper we do need this information, therefore
the choice of $D$ does not affect our conclusions.}.
This is the scope of the next section.

\section{The gas dynamics of modified shocks}

The velocity, density and thermodynamic properties of the fluid
can be determined by the usual conservation equations, including now
the pressure of the accelerated particles. We write
these equations between a point far upstream ($x=+\infty$), where the fluid
velocity is $u_0$ and the density is $\rho_0=m N_{gas,0}$, and a generic 
point where the fluid upstream velocity is $u_p$ (density $\rho_p$).
The index $p$ will denote quantities measured at the point where the
fluid velocity is $u_p$. We call this generic point $x_p$. 

The mass conservation implies:
\begin{equation}
\rho_0 u_0 = \rho_p u_p.
\label{eq:mass}
\end{equation}
Conservation of momentum reads:
\begin{equation}
\rho_0 u_0^2 + P_{g,0} = \rho_p u_p^2 + P_{g,p} + P_{CR,p},
\label{eq:pressure}
\end{equation}
where $P_{g,0}$ and $P_{g,1}$ are the gas pressures at the point 
$x=+\infty$ and $x=x_p$ respectively, and $P_{CR,p}$ is the pressure
in accelerated particles at the point $x_p$ (we used the symbol $CR$
to mean {\it cosmic rays}, to be interpreted here in a very broad sense).
In writing eqs. (\ref{eq:mass}) and (\ref{eq:pressure}) we implicitly 
assumed that the
average velocity $u_p$ as defined in eq. (\ref{eq:up}) coincides with the 
fluid velocity at the point where the particles with momentum $p$
turn around to recross the shock.

Our basic assumption, similar to that used in \cite{eich84a}, is that
the diffusion is $p$-dependent and that therefore particles with 
larger momenta move farther away from the shock than lower momentum
particles. This assumption is expected to describe what actually 
happens in the case of diffusion dependent on $p$. As a consequence, 
at each fixed $x_p$ only particles with momentum larger than $p$
are able to affect the fluid. Strictly speaking
the validity of the assumption depends on how strongly the diffusion 
coefficient depends on the momentum $p$, but the results should not
be critically affected by this assumption. Moreover, in case of strong
shocks there are arguments that suggest that the strong turbulence excited by
the shock should produce a B\"{o}hm diffusion coefficient, so that the
dependence $D(p)$ on $p$ should be at least linear. 

According to this assumption, only particles with momentum $\simgt p$ 
can reach the point $x=x_p$, therefore
\begin{equation}
P_{CR,p} = \frac{4\pi}{3} \int_{p}^{p_{max}} dp p^3 v(p) f(p),
\label{eq:CR}
\end{equation}
where $v(p)$ is the velocity of particles whose momentum is $p$, and 
$p_{max}$ is the maximum momentum achievable in the specific situation
under investigation. In realistic cases, $p_{max}$ is determined from 
geometry or from the duration of the shocked phase, or from the comparison
between the time scales of acceleration and losses. Here we consider it as a
parameter to be fixed {\it a priori}. From eq. (\ref{eq:pressure}) we
can see that there is a maximum distance, corresponding to 
the propagation of particles with momentum $p_{max}$ such that 
at larger distances the fluid is unaffected by the accelerated 
particles and $u_p=u_0$.

We will show later that for strongly modified shocks the integral in eq. 
(\ref{eq:CR}) is dominated by the region $p\sim p_{max}$. This improves
even more the validity of our approximation $P_{CR,p}=P_{CR}(>p)$.
This also suggests that different choices for the diffusion coefficient
$D(p)$ may affect the value of $p_{max}$, but at fixed $p_{max}$ the 
spectra of the accelerated particles should not be appreciably changed.

Assuming an adiabatic compression of the gas in the upstream region, 
we can write
\begin{equation} 
P_{g,p}=P_{g,0} \left(\frac{\rho_p}{\rho_0}\right)^{\gamma_g}=
P_{g,0} \left(\frac{u_0}{u_p}\right)^{\gamma_g},
\label{eq:Pgas}
\end{equation}
where we used the conservation of mass, eq. (\ref{eq:mass}). The 
gas pressure far upstream is $P_{g,0}=\rho_0 u_0^2/(\gamma_g M_0^2)$,
where $\gamma_g$ is the ratio of specific heats ($\gamma_g=5/3$ for an
ideal gas) and $M_0$ is the fluid Mach number far upstream. 
Note that eq. (\ref{eq:Pgas}) cannot be applied at the shock jump, where
the adiabaticity condition is clearly violated.

We can rewrite eq. (\ref{eq:pressure}) in a convenient way, by 
dividing it by $\rho_0 u_0^2$ and using the mass conservation. We
then obtain:
$$
1+\frac{1}{M_0^2 \gamma_g}=U+\frac{1}{M_0^2 \gamma_g} U^{-\gamma_g} + 
\frac{4\pi}{3} \frac{1}{\rho_0 u_0^2} \int_{p}^{p_{max}} dp' p'^3 v(p') 
f_0(p')=
$$
\begin{equation}
U+\frac{1}{M_0^2 \gamma_g} U^{-\gamma_g}+
\frac{N_{inj} q_s}{\rho_0 u_0^2 p_{inj}^3} 
\int_p^{p_{max}} dp' p'^3 v(p') \exp
\left\{-\int_{p_{inj}}^{p'} \frac{dp''}{p''} \left[
\frac{3 u_{p''}}{u_{p''}-u_2} + \frac{1}{u_{p''}-u_2}\frac{du_{p''}}{d\ln p''}
\right]\right\},
\end{equation}
where we have explicitely written the distribution function 
$f(p)$ as a power law with local slope $Q(U)$, and we put 
$U=u_p/u_0$. 

Differentiating the previous equation with respect to $p$ we
obtain:
\begin{equation}
\frac{d U}{d p} \left[1 - \frac{1}{M_0^2} U^{-(\gamma_g+1)}\right]=
\frac{1}{3}\frac{N_{inj}q_s}{\rho_0 u_0^2} 
\left(\frac{p}{p_{inj}}\right)^3 v(p)  \exp
\left\{-\int_{p_{inj}}^p \frac{dp}{p} \left[
\frac{3 u_p}{u_p-u_2} + \frac{1}{u_p-u_2}\frac{du_p}{d\ln p}\right]
\right\}
\label{eq:diff0}
\end{equation}
or 

$$
\frac{d U}{d \ln p} \left[1 - \frac{1}{M_0^2} U^{-(\gamma_g+1)}\right]=
$$
\begin{equation}
\frac{1}{3}\frac{N_{inj}q_s}{\rho_0 u_0^2} 
v(p) p_{inj} \exp
\left\{ 4\ln\left(\frac{p}{p_{inj}}\right)-
\int_{p_{inj}}^p \frac{dp}{p} \left[
\frac{3 u_p}{u_p-u_2} + \frac{1}{u_p-u_2}\frac{du_p}{d\ln p}\right]\right\}.
\label{eq:diff}
\end{equation}
Note that the velocity $u_p$ changes as a consequence of the
pressure added by non-thermal particles, therefore the function
$U(p)$ must be a monotonically increasing function of the
particle momentum. Since $U(p)$ and $dU/d\ln p$ are always non-zero, 
we can calculate their logarithm, so that eq. (\ref{eq:diff}) becomes:
$$
\ln {\cal D}U + \ln \left[1 - \frac{1}{M_0^2} U^{-(\gamma_g+1)}\right]
$$
\begin{equation}
=\ln \left[ \frac{1}{3}\frac{N_{inj}q_s}{\rho_0 u_0^2} 
v(p) p_{inj}\right] + 4\ln\left(\frac{p}{p_{inj}}\right) -
\int_{p_{inj}}^p \frac{dp}{p} \left[
\frac{3 R_{tot} U}{R_{tot} U - 1} + \frac{R_{tot}}{R_{tot} U - 1} {\cal D}U
\right],
\end{equation}
where we put ${\cal D}U=dU/d\ln p$.
The following equality is easily demonstrated:
\begin{equation}
\int_{p_{inj}}^p \frac{dp}{p} \frac{R_{tot}}{R_{tot} U - 1} {\cal D}U
=\ln\left(\frac{R_{tot}U-1}{R_{sub}-1}\right),
\end{equation}
so that the equation for ${\cal D}U$ becomes:
$$
\ln {\cal D}U + \ln \left[1 - \frac{1}{M_0^2} U^{-(\gamma_g+1)}\right]
$$
\begin{equation}
=\ln \left[ \frac{1}{3}\frac{N_{inj}q_s}{\rho_0 u_0^2} 
v(p) p_{inj}\right] + 4\ln\left(\frac{p}{p_{inj}}\right) -
\ln\left(\frac{R_{tot}U-1}{R_{sub}-1}\right)-
\int_{p_{inj}}^p \frac{dp}{p} \frac{3 R_{tot} U}{R_{tot} U - 1} 
\label{eq:nobel}
\end{equation}
Solving this differential equation provides $U(p)$ and therefore the
spectrum of accelerated particles, through eq. (\ref{eq:implicit}).

The operative procedure for the calculation of the spectrum of accelerated
particles is simple: we fix the boundary condition at $p=p_{inj}$ such that 
$U(p_{inj})=u_1/u_0$ for some value of $u_1$ (fluid velocity at $x=0^+$). 
The evolution of $U$ as a function of $p$ is determined by 
eq. (\ref{eq:nobel}). 
The physical solution must have $U(p_{max})=1$ because at $p\simgt p_{max}$
there are no accelerated particles to contribute any pressure. There is a 
unique value of $u_1$ for which the fluid velocity at the prescribed 
maximum momentum $p_{max}$ is $u_{p_{max}}=u_0$ (or equivalently 
$U(p_{max})=1$). Finding this value of
$u_1$ completely solves the problem, since eq. (\ref{eq:nobel}) provides
$U(p)$ and therefore the spectrum of accelerated particles, calculated
according to eq. (\ref{eq:spec}). Conservation of energy can be 
easily checked.

\begin{figure}[thb]
 \begin{center}
  \mbox{\epsfig{file=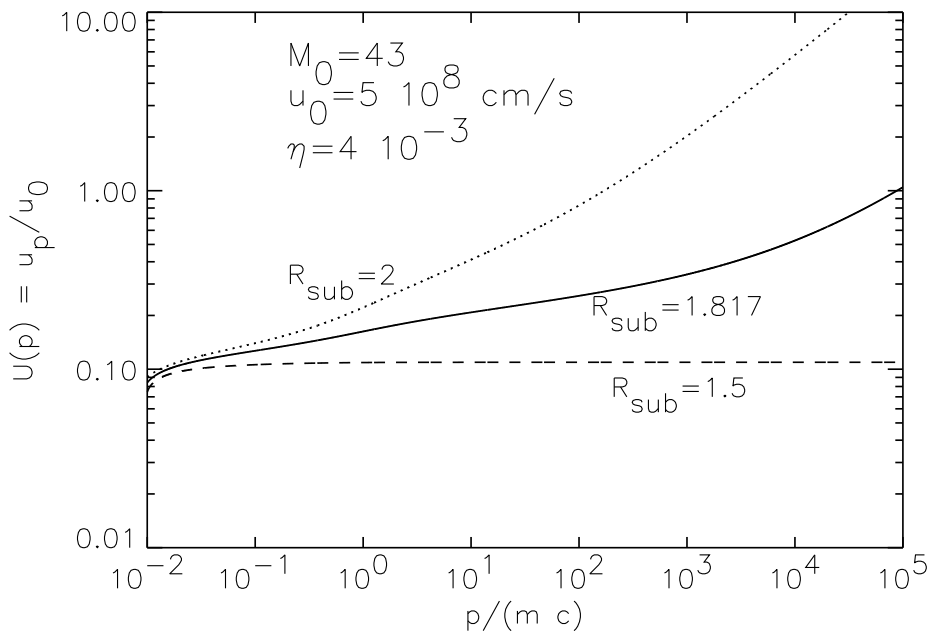,width=13.cm}}
  \caption{\em {$U(p)$ as a function of momentum, for different
values of the compression factor at the subshock. The fluid dynamics
must be unaffected at $p=p_{max}$ which implies that $U(p_{max})=1$. 
This condition determines the right value of $R_{sub}$.
}}
 \end{center}
\end{figure}

An illustration of this procedure is presented in fig. 1
where we considered
the following special set of parameters: the Mach number far upstream
is $M_0=43$, and the gas temperature is $T_0=10^6$ K, corresponding to 
$u_0=5\times 10^8$ cm/s. The parameter $\eta$ is taken to be $4\times 10^{-3}$,
and the injection of particles occurs at $p_{inj}=10^{-2} mc$, where
$m$ is the mass of the accelerated particles. In this paper we assume that
the accelerated particles are protons. The 
maximum momentum is $p_{max}=10^5 mc$. The three curves which 
are plotted correspond to the function $U(p)$ for $R_{sub}=2$ (dotted
line), $R_{sub}=1.5$ (dashed line) and $R_{sub}=1.817$ (solid line). 
It is immediately evident that only for $R_{sub}=1.817$ the function
$U(p)$ reaches unity at $p=p_{max}$. Our method then 
provides the value of $R_{sub}$ and consequently the values of the other 
parameters.

Fig. 1 is very useful for understanding the physical meaning of the 
local velocity $u_p$. Particles with large momenta {\it feel} 
compression factors $u_p/u_2$ which are larger than those felt
by low momentum particles. Large compression factors correspond to
locally flatter spectra, so that the spectrum of accelerated particles 
is expected to become flatter at large momenta. We define the total 
compression factor $R_{tot}=u_0/u_2$.  
Note that the compression factor at the gas subshock is now
$R_{sub}\ll 4$, the value expected for a strong shock in the
test particle regime.

\section{Results}

The calculations illustrated in the previous section are here tested
versus previous models and numerical simulations. In particular we 
compare our results with the predictions of the simple approach 
presented in \cite{simple}, that we briefly summarize. In \cite{simple}
the spectrum of accelerated particles has a prescribed shape, made of
three power laws, in the low ($p_{inj}\leq p\leq mc$), intermediate
($mc \leq p\leq 10^{-2} p_{max}$) and high 
($10^{-2} p_{max}\leq p\leq p_{max}$) energy regimes. The slope in the
three regions is then calculated by requiring mass, momentum and energy 
conservation. 

We first check that our model reproduces the results in the linear
regime, where the test particle approximation can be adopted, and then
we study the transition to the non-linear regime. 
For this purpose, we consider a shock with Mach number $M_0=5$,
with a gas temperature $T_0=10^8$ K. We choose $p_{inj}=10^{-2} mc$ and
$p_{max}=10^5 mc$, and we study the result for different values of $\eta$,
as plotted in Fig. 2. The solid line is obtained for $\eta=10^{-5}$; our
model gives for this case $R_{sub}=R_{tot}=3.57$, which is exactly the 
value obtained from test particle theory: $R_{sub}=(8M_0/3)/((2/3)M_0+2)$,
for $\gamma_g=5/3$. The test particle approximation would provide the same
compression factor for any value of $\eta$. Our results for $\eta=10^{-3}$ and 
$\eta=10^{-2}$ are plotted in Fig. 2 as dotted and dashed lines respectively.
These two cases result in $(R_{sub},R_{tot})=(3.45,3.83)$ for $\eta=10^{-3}$
and $(R_{sub},R_{tot})=(2.85,4.49)$ for $\eta=10^{-2}$. The transition 
from unmodified shocks to strongly modified shocks is evident.  

\begin{figure}[thb]
 \begin{center}
  \mbox{\epsfig{file=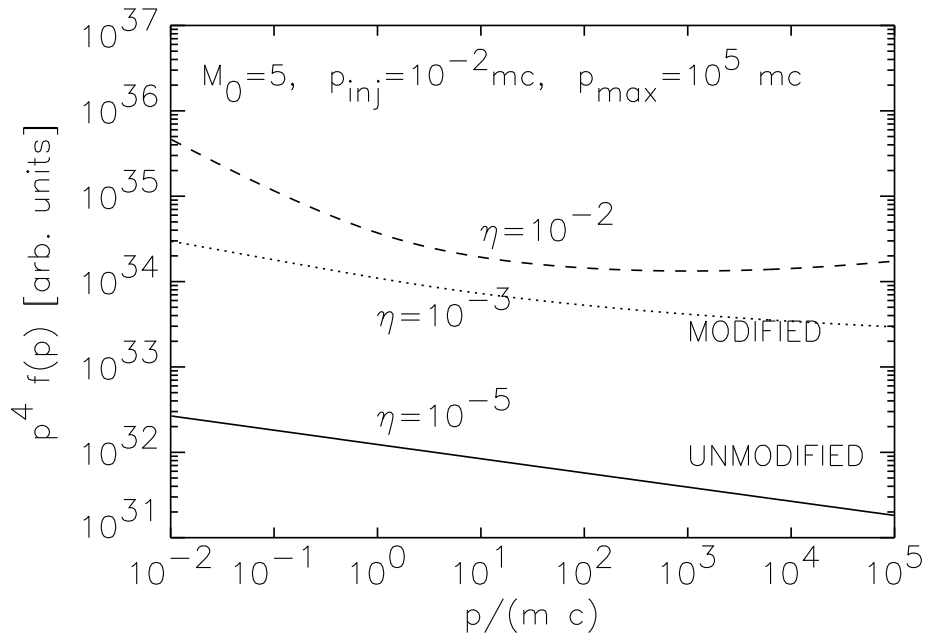,width=13.cm}}
  \caption{\em {Spectra of accelerated particles for low Mach number ($M_0=5$).
Increasing the value of $\eta$ determines the transition from an unmodified
shock (solid line, $\eta=10^{-5}$) and a modified shock (dashed and dotted
lines, with $\eta=10^{-2}$ and $\eta=10^{-3}$ respectively).
}}
 \end{center}
\end{figure}

In all the cases the common feature of the modified shocks is a steepening
of the spectrum (in comparison with the linear result) at low energy
and a flattening at high energies. From the phenomenological point of 
view this 
can be of paramount importance since it may change the spectral features
of the secondary radiation produced by the interactions of the accelerated
particles.

The more interesting case is that of strongly modified shocks, where we
expect the pressure in accelerated particles to become comparable with
the kinetic energy of the upstream fluid. Intuitively this is mainly the
case for high Mach number shocks, though we also show that there are 
exceptions.
We can compare our results with those of the model in Ref. \cite{simple}.

Let us consider the case $M_0=43$, $u_0=5\times 10^8$ cm/s, 
$p_{inj}=10^{-2} mc$, and let us study the resulting spectra for  
$\eta=10^{-3}$ and different values of $p_{max}$. The 
results are plotted in Fig. 3. 

\begin{figure}[thb]
 \begin{center}
  \mbox{\epsfig{file=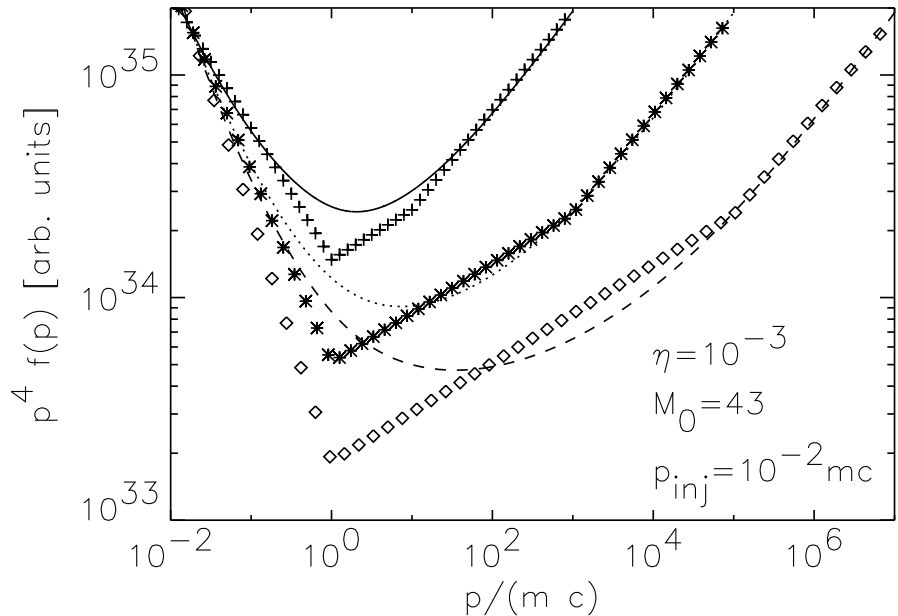,width=13.cm}}
  \caption{\em {Comparison between the prediction of our model (lines) and
those of Ref. \cite{simple} (symbols). The three sets of curves are obtained
for $p_{max}=10^3 mc$ (solid line and crosses), $p_{max}=10^5 mc$
(dotted line and stars) and $p_{max}=10^7 mc$ (dashed line and diamonds).
}}
 \end{center}
\end{figure}

The results of \cite{simple} are plotted in the form of crosses 
($p_{max}=10^{3}mc$), stars ($p_{max}=10^{5}mc$) and diamonds
($p_{max}=10^{7}mc$). The corresponding continuous lines are the
results of our model.

Some comments are in order: the slope of the spectrum of accelerated
particles predicted in our model at $p\sim p_{max}$ is approximately
equal to that obtained in \cite{simple}.
This is not surprising since the model in \cite{simple} is based on 
a three power law approximation, and the slope of the spectrum at the highest
energies is calculated using an asymptotic expression derived from 
eq. (\ref{eq:implicit}). At low values of $p_{max}$ our results are in very 
good agreement with the general features of the solutions in \cite{simple}.
At increasingly larger values of $p_{max}$ the agreement is not extremely 
good but still reasonable, if one takes into account that the following 
three assumptions were made in \cite{simple}: 1) the velocity of the
particles is assumed to be $p/m$ for $p\leq mc$ and equal to $c$ for
$p\geq mc$; 2) the momentum at which there is a change in slope 
at low energy is forced to be at $p=mc$; 3) the position of the point 
where there is the change in slope at intermediate energies is forced to 
be at $p=10^{-2} p_{max}$.
In our model all the three assumptions are released and the spectra are 
smooth. This is the reason for the slightly different position of the dip 
in our spectra ($p^4 f(p)$) when compared with those derived according 
to \cite{simple}.
Similar conclusions hold for different values of the parameters.

A more interesting comparison, to test the effectiveness of our model
is that with the results of numerical simulations. Since a similar comparison
was carried out in \cite{simple} to test that model, we consider here the same
situation, so that a full cross-check is possible. The case we consider is that
of a shock with Mach number $M_0=128$ and a fluid temperature of $T=10^6$ K. 
The results provided by a numerical simulation are plotted in Fig. 5 of Ref.
\cite{simple} and reproduced in our Fig. 4 as a dashed line. From the simulation
we can extract the values of some parameters:
the injection momentum is $p_{inj}=7\times 10^{-3} mc$, while the maximum
momentum is $p_{max}=10^5 mc$. Note that the simulation does not provide these
parameters in a clear way, because there the thermal and non-thermal particles
are treated in the same way, therefore some approximation is involved in 
deriving these numerical values. The best fit of the model in \cite{simple}
to the results of the simulation implies $R_{sub}=2.68$ and $R_{tot}=52$
using $\eta=5\times 10^{-3}$. Using the same value of the injection 
efficiency $\eta$, our model predicts 
$R_{sub}=2.365$ and $R_{tot}=51.9$. The spectra for the model in \cite{simple}
and for our model are plotted in Fig. 4 as a solid light broken line, and as a 
thick solid line respectively. The improvement provided by our model is
evident (note that we only used the values of $\eta$ and $p_{inj}$
derived in
\cite{simple}, though an even better agreement can be found by slightly
changing these values, that, as we stressed above, are
not clearly determined by the simulation).

\begin{figure}[thb]
 \begin{center}
  \mbox{\epsfig{file=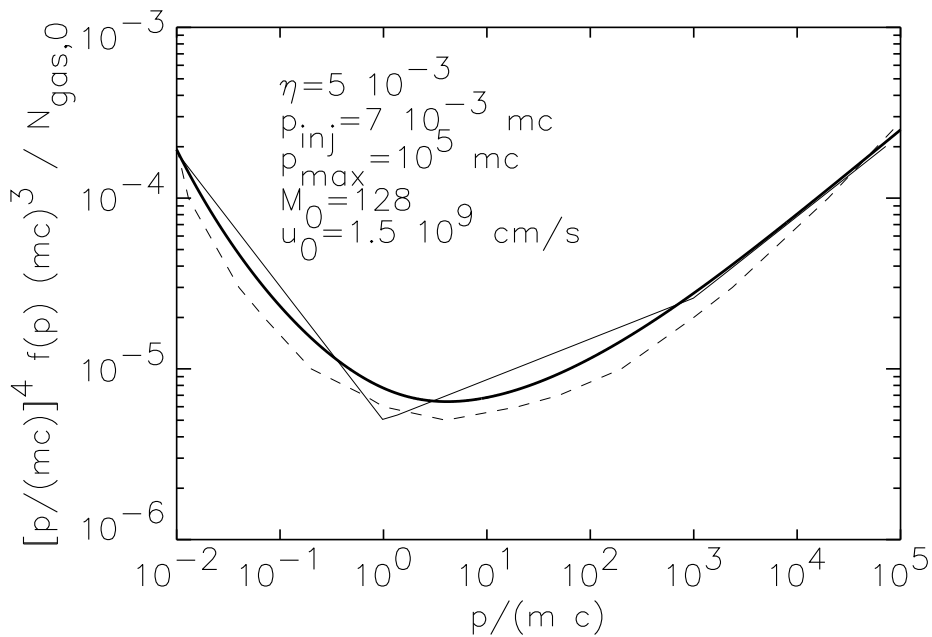,width=13.cm}}
  \caption{\em {Comparison between the predictions of our model (thick solid
line) and the results of simulations (dashed line) and the approximation in
Ref. \cite{simple} (solid light line).
}}
 \end{center}
\end{figure}

A basic issue raised in \cite{simple} is that of the accurate description of
the slope of the spectrum of accelerated particles at $p\sim p_{max}$. 
The importance of this point is due to the flat spectral shape at high
momenta, which gives the main contribution to the total energy budget
in accelerated particles, for strongly modified shocks.
We show our prediction for the local slope in Fig. 5, as a function 
of the momentum. The most noticeable feature of this figure is the
slope $\sim 3.5$ at $p\sim p_{max}$, very close to that predicted by
numerical simulations.

\begin{figure}[thb]
 \begin{center}
  \mbox{\epsfig{file=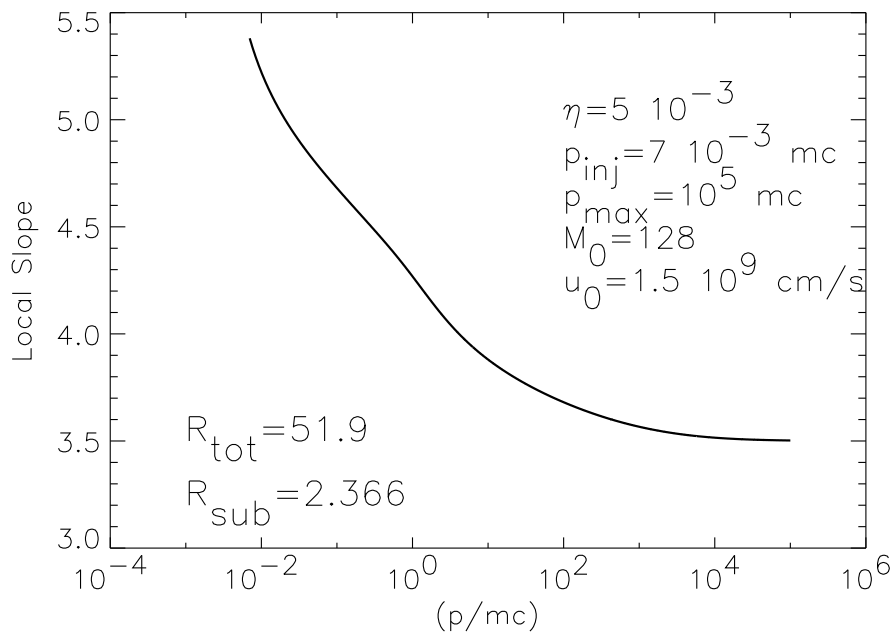,width=13.cm}}
  \caption{\em {Slope of the spectrum of accelerated particles according
with our model, with the parameters used to obtain the curves in Fig. 4.
}}
 \end{center}
\end{figure}

One comment is in order concerning the injection: in all the cases
considered above,
the injection momentum and $\eta$, the injection efficiency, are chosen 
independently. In a realistic case (for instance in the simulation) the
two parameters would be actually related to each other. For instance, if the 
particles are extracted from the thermal distribution, then one can write 
in a very general way, that the injection momentum is proportional to the 
sound speed downstream ($c_{s,2}$), times the particle mass, 
$p_{inj}=\xi m c_{s,2}$. The parameter $\xi$
must be large enough that the particles can {\it feel} the thickness
of the shock, determined either by the interaction pathlength of the 
particles, or by the Larmor radius of the thermal protons in the local 
magnetic field.
Hence, $\xi$ must be at least a few, and indeed it is usually assumed to be 
in the range $\xi\sim 4-10$. If the expression $p_{inj}=\xi m c_{s,2}$ 
is adopted, and the injection is assumed to occur from the thermal
distribution, then the parameter $\eta$ is no longer free, and can be
determined self-consistently. However it 
is wise to keep in mind that in this case $\eta$ would change wildly for
small changes in $\xi$, due to the exponential suppression in the 
distribution function at momenta larger than the thermal average.

We continue now the analysis of the predictions of our model and the
comparison with the results of Ref. \cite{simple}.
An important issue is that of determining where the transition 
from unmodified to strongly modified shock occurs, as a function of
the parameters of the calculation. 

In Fig. 6 we plot the compression ratios (total and at the subshock) for 
different values of the maximum momentum of the accelerated particles.
The solid lines represent the results of our model, compared with the
values predicted by the simple approach in \cite{simple} (dashed lines).
There is a good agreement between the two approaches. For $\eta=10^{-3}$
the shock becomes strongly modified already for $p_{max}$ a few times larger
than $mc$, confirming that even relatively low injection efficiencies cause
the shock to be affected by the backreaction of the accelerated particles.

\begin{figure}[thb]
 \begin{center}
  \mbox{\epsfig{file=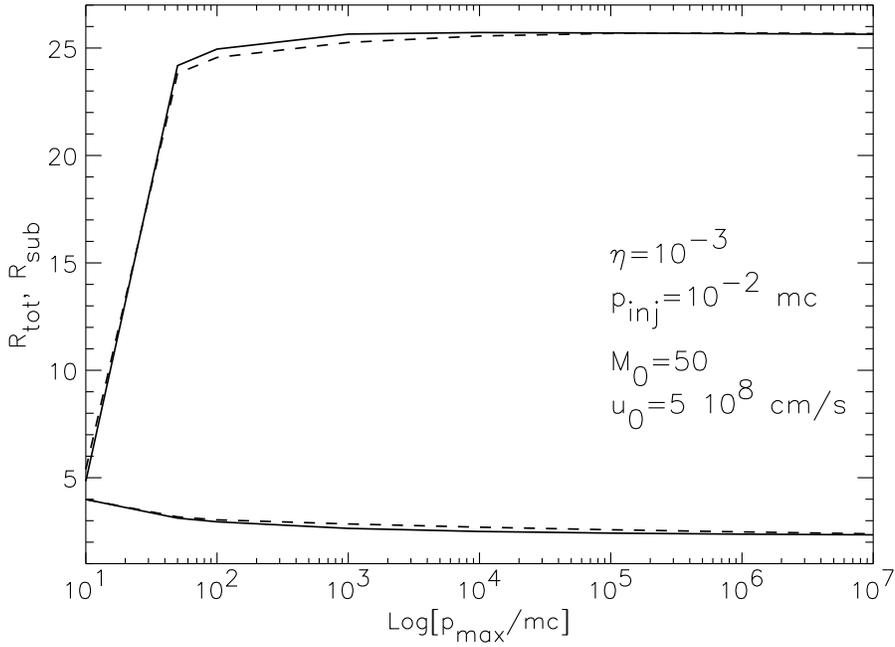,width=13.cm}}
  \caption{\em {Compression factors according with our model
(solid lines) and the simple model (dashed lines), as functions 
of $p_{max}$, for the values of the parameters reported in the figure.
The upper curves represent $R_{tot}$ and the lower curves represent $R_{sub}$.
}}
 \end{center}
\end{figure}

In Fig. 7 the same compression factors are plotted versus the minimum 
(injection) momentum $p_{inj}$. Here some differences between our model and
that in Ref. \cite{simple} are visible. The results can be 
interpreted as an evidence that our model predicts the onset of the 
modified shock phase, at slightly larger values of $p_{inj}$ compared
to the simple model.
 
\begin{figure}[thb]
 \begin{center}
  \mbox{\epsfig{file=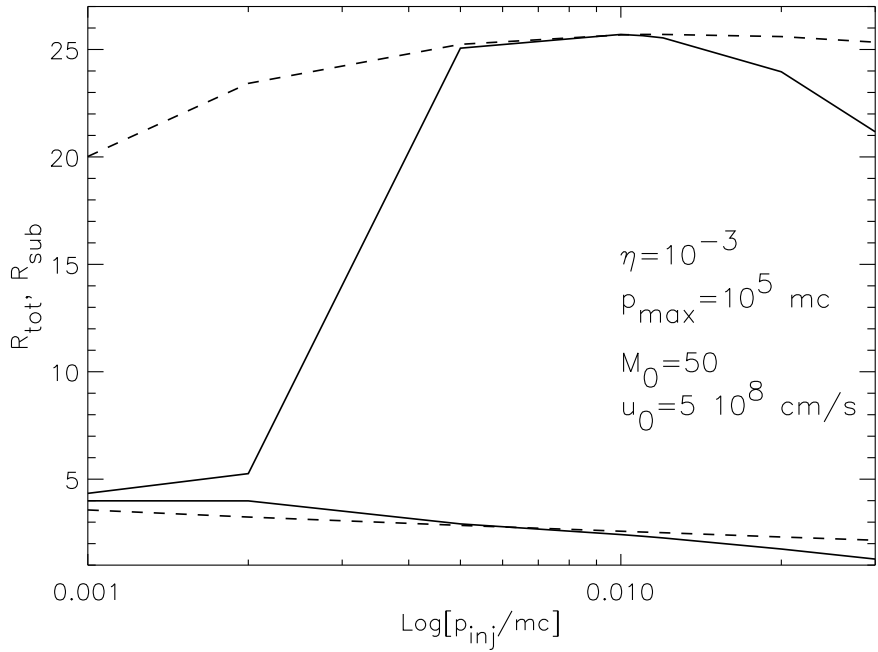,width=13.cm}}
  \caption{\em {Compression factors according with our model
(solid lines) and the simple model (dashed lines) as functions
of the injection momentum. The upper curves represent $R_{tot}$ and the 
lower curves represent $R_{sub}$.
}}
 \end{center}
\end{figure}

The results of a similar study are plotted in Fig. 8, where the dependence of 
$R_{tot}$ and $R_{sub}$ is investigated versus the injection efficiency $\eta$.
Again, our model predicts that the shock starts to be modified 
at values of $\eta$ slightly larger than for the simplified model.
In particular for the values of the parameters reported in Fig. 8,
we obtain that the shock starts to be modified at $\eta\simgt 10^{-4}$,
while the simple model gives a modified shock for 
$\eta\simgt 3\times 10^{-5}$. 
\begin{figure}[thb]
 \begin{center}
  \mbox{\epsfig{file=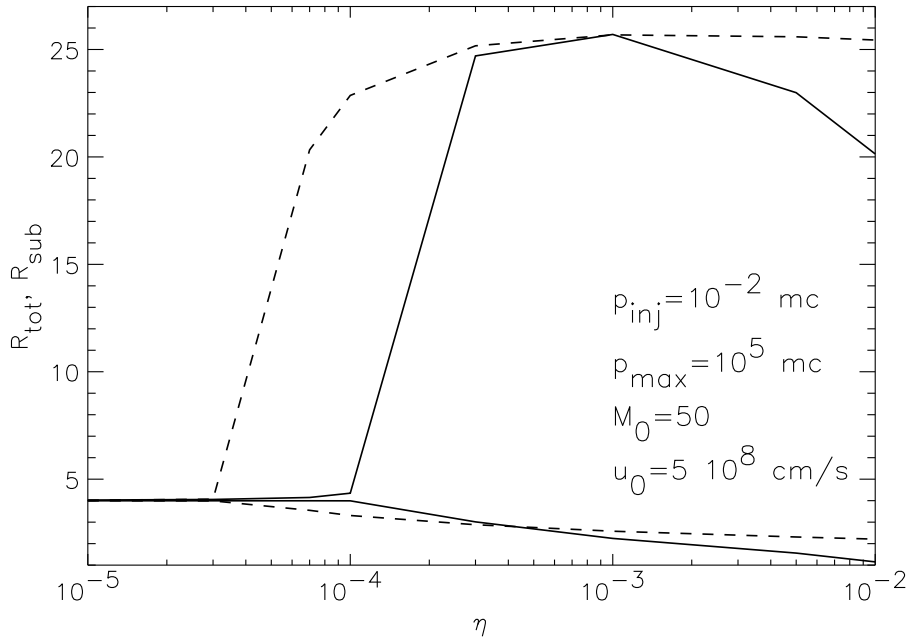,width=13.cm}}
  \caption{\em {Compression factors according with our model
(solid lines) and the simple model (dashed lines) as functions of
the injection efficiency $\eta$.  The upper curves represent $R_{tot}$ and the 
lower curves represent $R_{sub}$.
}}
 \end{center}
\end{figure}
Finally in Fig. 9 we investigate the dependence of $R_{tot}$ and $R_{sub}$
on the Mach number of the fluid at infinity (upstream). We specialize
our prediction to the case $\eta=10^{-3}$, $p_{inj}=10^{-2}mc$ and
$p_{max}=10^{5}mc$, but clearly similar plots can be produced for
different regions of the parameter space. The results of our model are in 
very good agreement with the results of \cite{simple}. Our model however
predicts a slightly lower value of the critical Mach number $M_{cr}$, above
which the shock is no longer modified, so that $R_{tot}$ and $R_{sub}$ 
settle down at the usual (linear) value of $\sim 4$. As found in \cite{simple},
the relation between $R_{tot}$ and $M_0$ is $R_{tot}\propto M_0^{3/4}$,
for $M_0\leq M_{cr}$.

\begin{figure}[thb]
 \begin{center}
  \mbox{\epsfig{file=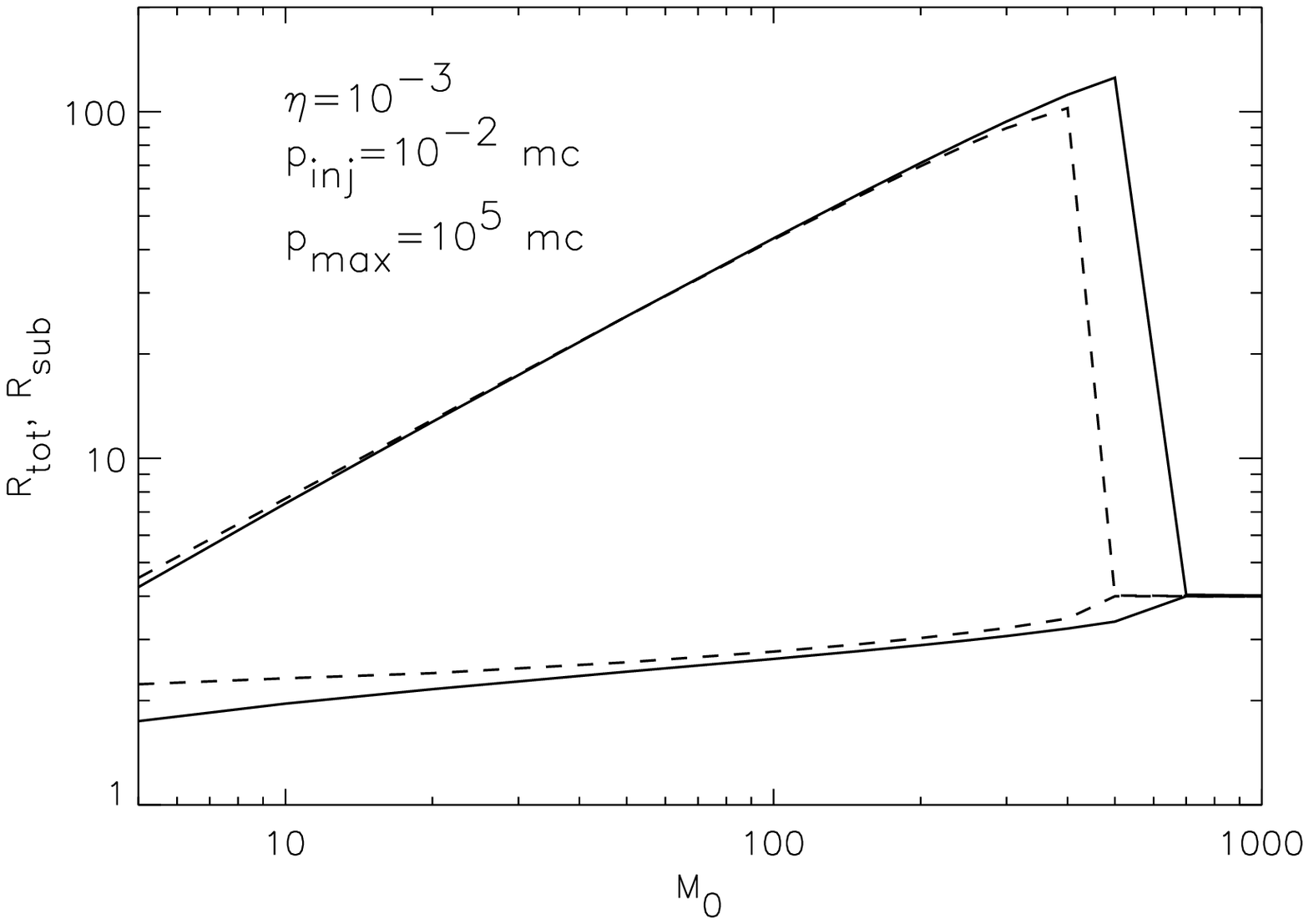,width=13.cm}}
  \caption{\em {Compression factors according with our model
(solid lines) and the simple model (dashed lines) as functions 
of the Mach number $M_0$.  The upper curves represent $R_{tot}$ and the 
lower curves represent $R_{sub}$.
}}
 \end{center}
\end{figure}

\section{Discussion and Conclusions}

We proposed a semi-analytical method to calculate the spectrum of 
particles accelerated at shocks, and the velocity profile of the shocked 
fluid, including the non-linear effects
due to the backreaction of the non-thermal particles on the dynamics
of the shocked fluid. 

The approach presented in this paper provides results that are in 
good agreement with the output of numerical simulations on shock 
acceleration and also with the results of a previous model aimed to 
a simple description of the non-linear effects \cite{simple}.
Compared with the latter, our model provides a better fit to numerical 
results, and is not based on {\it a priori} assumptions on the
spectrum of accelerated particles in some energy ranges.

We confirm the main results on shock acceleration, known 
from simple models and from simulations: {\it i)} the backreaction of
the accelerated particles is important, even in those cases
in which a small fraction of the particles 
injected at the shock is actually accelerated to suprathermal 
energies. For $\eta\simgt 10^{-4}$ the shock becomes modified by
the non-thermal pressure and both the velocity field of the shocked
fluid and the spectrum of the accelerated particles are affected.
The minimum value of $\eta$ for which the shock is modified in our model is
slightly larger than that predicted in \cite{simple}.
{\it ii)} For linear shocks the maximum compression factor that can 
be achieved is $4$ (for $\gamma_g=5/3$) which corresponds to spectra
$\propto p^{-4}$. This result is obtained for $M_0\to \infty$. 
When the backreaction becomes relevant, the structure of the shocked
fluid is changed into a smooth decrease of the fluid speed from $u_0$
at infinity (effectively at some distance $D_{max}$) to $u_1$ at the position
of the gas (ordinary) subshock. While at the subshock it is still true
that the maximum compression factor is $4$, the overall compression
factor ($R_{tot}$) between downstream and $D_{max}$ can be arbitrarily
large. {\it iii)} The large total compression factors for strongly modified
shocks result in a flattening of the spectra of accelerated particles
at high energy. The slope there tends to $\sim 3.5$ (flatter than $4$).
The slope of the spectrum at low energy is determined by the compression
factor at the subshock, and is usually steeper than the slope predicted
in linear theory.
{\it iv)} In general, increasing the Mach number corresponds to an
increasingly more modified shock, up to a critical Mach number $M_{cr}$.
At $M>M_{cr}$, the total and subshock compression factors both converge
to the linear value of $4$ and the shock behaves as an ordinary strong 
shock (it is no longer an efficient accelerator).

Although the calculations reported in the paper refer to the case of 
a plane (one dimensional) shock and the effects of waves (Alfv\`en 
heating) have not been included, these changes are almost straightforward. 
In fact,  for spherical shocks, the expression for the distribution function
$f_0(p)$ remains unchanged [eq. (\ref{eq:implicit})], although the
definition of $u_p$ is formally different (it reflects the
geometry). All the rest of the calculation remains unaffected and most of
the results still hold. The case of spherical symmetry would be relevant
for shocks related to supernova remnants, as discussed in \cite{berezhko96}.

The introduction of the heating due to damping of Alfv\'en waves 
implies a change in the relation between $P_{g,p}$ and $P_{g,0}$ 
but this change can be easily written in a way which is useful for
our purposes. For instance, generalizing a discussion in \cite{simple},
in the approximation of large Alfv\'en Mach number ($M_{A}=u_0/v_A$, 
where $v_A$ is the Alfv\'en speed, assumed constant in all the fluid)
we can write
\begin{equation}
\frac{P_{g,p}}{P_{g,0}}\simeq \left(\frac{\rho_p}{\rho_0}\right)^{\gamma_g}
\left\{1+(\gamma_g-1) \frac{M_0^2}{M_A}\left[1-
\left(\frac{\rho_0}{\rho_p}\right)^{\gamma_g}\right]\right\}.
\label{eq:alfven}
\end{equation}
Clearly the effects of Alfv\'en heating can be neglected as long as
$M_0^2\ll M_A$.
Introducing eq. (\ref{eq:alfven}) [instead of eq. (\ref{eq:Pgas})]
in eq. (\ref{eq:pressure}), it is easy to derive an equation similar to
eq. (\ref{eq:nobel}), that can be solved for $U(p)$. The spectrum of
accelerated particles is then obtained using the same procedure 
illustrated above. The basic effect of the Alfv\'en heating is to
reduce the total compression factors, and make the spectra at high energy
slightly steeper than found in \S4.

We conclude by stressing that the non-linear effects discussed here
appear to be relevant even for a small injection efficiency, and their
phenomenological consequences can be critically important. For instance,
as an application of the simple model presented in \cite{simple}, in Ref.
\cite{simple_appl} the spectra of secondary radio and gamma radiation 
in supernova remnants were calculated. The differences with respect to the 
results obtained by adopting the test particle approximation are impressive.

In this prospective, the semi-analytical model presented here, being
of easy use and having an immediate physical interpretation, provides
the suitable tool to estimate the phenomenological consequences of
non-linearity in shock acceleration, in the cases where it is unpractical
to have access to numerical simulations.

{\bf Acknowledgments} 
This work was supported by the DOE and the NASA grant NAG 5-7092 
at Fermilab.

\end{document}